# Accurate prediction of gene expression by integration of DNA sequence statistics with detailed modeling of transcription regulation

Jose M. G. Vilar[1,2]

[1] Biophysics Unit (CSIC-UPV/EHU) and Department of Biochemistry and Molecular Biology, University of the Basque Country, P.O. Box 644, 48080 Bilbao, Spain

[2] IKERBASQUE, Basque Foundation for Science, 48011 Bilbao, Spain

## Abstract

Gene regulation involves a hierarchy of events that extend from specific protein-DNA interactions to the combinatorial assembly of nucleoprotein complexes. The effects of DNA sequence on these processes have typically been studied based either on its quantitative connection with single-domain binding free energies or on empirical rules that combine different DNA motifs to predict gene expression trends on a genomic scale. The middle-point approach that quantitatively bridges these two extremes, however, remains largely unexplored. Here, we provide an integrated approach to accurately predict gene expression from statistical sequence information in combination with detailed biophysical modeling of transcription regulation by multidomain binding on multiple DNA sites. For the regulation of the prototypical *lac* operon, this approach predicts within 0.3-fold accuracy transcriptional activity over a 10,000-fold range from DNA sequence statistics for different intracellular conditions.

# Introduction

In a now classic paper proposing the *lac* operon model, Jacob and Monod put forward the very basic principles of gene regulation (1). They reasoned that there are molecules that bind to specific sites in nucleic acids to control whether or not genes are expressed. Since then, a major challenge in biology has been to understand how site-specific regulatory factors function and the effects that they have in gene regulation. Thus, over the last decades, there have been a large effort to produce reliable and efficient computer algorithms for the analysis and prediction of DNA binding sites (2).

These algorithms have reached an extraordinary ability at predicting with high accuracy how proteins bind single sites (3, 4). At the same time, use of these highly accurate models to predict where additional binding sites might occur typically finds a wealth of sites that are not physiologically relevant (2). A rule of thumb to predict actual binding is that relevant sites often appear close to each other to act cooperatively (5). Clever refinement of this idea has led to heuristic approaches that have proved very successful at predicting the main gene expression trends on a genomic scale (6-12). The middle ground between detailed single site and broad genomic predictions, however, still remains largely unexplored.

Here, we develop a quantitative framework that accurately integrates sequence statistics with a biophysical model for multidomain binding on non-adjacent DNA sites using as a prototype system the *lac* operon. This choice is motivated by two key features of the *lac* operon.

First, the very simple, yet extremely powerful, original idea of the *lac* repressor preventing transcription upon binding to the operator DNA in the promoter region has kept evolving over the years to uncover a highly sophisticated mechanism that goes beyond simple binding events (13). It incorporates now an activator and two additional binding sites for the repressor outside the promoter region. These two additional sites are orders of magnitude weaker than the main one and by themselves do not affect transcription substantially. In combination with the main one, however, they can increase repression of transcription by almost a factor 100 (14, 15).

Second, there is extremely detailed information about the *lac* operon that offers the possibility of considering the actual mode of binding. This point is important because the precise sequence has been shaped by evolution through the actual biophysical mechanism. The available information includes detailed quantitative models of how the *lac* repressor binds to two sites simultaneously (16, 17) and to the three sites for the repressor together with the effects of the Catabolite Activator Protein (CAP) (18, 19). The values of the molecular and cellular parameters needed by the models are also available, including the *in vivo* free energy of binding, the energetic costs of bending and twisting DNA upon two-site binding, and the effective transcription rate as a function of the binding state of the repressor (13, 20).

Therefore, the *lac* operon provides an efficient platform to accurately test multisite models. In this classical example, without considering the two additional sites, no matter how good the single site model is, it would be off by a factor of almost 100.

The focus here is to provide an avenue to extend traditional biophysical single-domain-binding models (21-23) to incorporate the details of multidomain binding, which are inherently different from those of single-domain binding of multiple transcription factors. The traditional approach considers the interaction of a transcription factor, $TF$, with a DNA site, $S1$, as a binding reaction of the type $TF + S1 \Leftrightarrow TF \cdot S1$. The strength of the binding is typically assessed through position weight matrix (PWM) scores, which are directly related to the binding energy of the DNA–protein interaction (3, 24). The extension to multidomain transcription factors in the presence of additional binding sites, denoted $S2$ and $S3$, has to consider also reactions of the type $TF \cdot S1 + S2 \Leftrightarrow S2 \cdot TF \cdot S1$ and $S2 \cdot TF \cdot S1 + S3 \Leftrightarrow S3 \cdot TF \cdot S1 + S2$. These more complex reactions account for binding of one domain of the TF while its other domain is still bound to DNA and usually involve looping the DNA between each pair of simultaneously bound sites.

The multisite approach is explicitly implemented by considering first the three *lac* operators as DNA signals. They are used to construct a probabilistic model that provides PWM scores for binding of a *lac* repressor domain to these and similar mutated sequences. The scores are subsequently linked parametrically to binding free energies and incorporated directly into a detailed biophysical model of transcription regulation that takes into account multidomain binding to multiple binding sites. The model considers a decomposition of the free energy of the protein-DNA complex into different modular contributions. The link between scores and free energies is calibrated by fitting the model to a subset of experimental transcription data. The calibrated model is then tested with different sets of data (Figure 1).

## Methods

### From sequence to score

The PWM method is used to describe repressor-operator binding (3, 24). It assigns a score $S$ to the sequence $X = x_1 x_2 ... x_w$ according to

$$S = \sum_{i=1}^{w} \ln \frac{p_{xi}}{q_x}, \tag{1}$$

where $p_{xi}$ is the estimated probability of having the nucleotide $x$ at position $i$ of the binding site and $q_x$ is the background frequency of that nucleotide. Taking into account small sample size, $p_{xi}$ is estimated from the observed positional frequency as

$$p_{xi} = \frac{n_{xi} + 1}{N + 4}, \tag{2}$$

where $n_{xi}$ is the number of sites having a nucleotide $x$ at position $i$ and $N$ is the total number of sites in the training set. In our case, we have only three sequences in the training set corresponding to the three operators.

### From score to free energy

We assume a linear relationship to transform the score $S$ of each sequence into the interaction free energies, $e$, between the *lac* repressor domain and the DNA site:

$$e = aS + b, \tag{3}$$

where $a$ and $b$ are constants to be inferred from experiments. With this linear assumption, $a$ selects the energy units and $b$ the reference zero of energy.

## Multidomain binding

The *lac* repressor is a tetramer consisting of two dimeric DNA binding domains. Multidomain binding is taken into account by decomposing the free energy of the protein-DNA complex into different modular contributions, including positional, interaction, and conformational free energies (19, 25).

The positional free energy, $p$, accounts for the cost of bringing the *lac* repressor to its DNA binding site. Its dependence on the repressor concentration, $n$, is given by $p = p^\circ - RT \ln n$, where $p^\circ$ is the positional free energy at 1M. Interaction free energies, $e$, arise from the physical contact between a binding domain and DNA site. Thus, when only a single domain is involved, the free energy of binding is given by $\Delta G = e + p$. For two domains, denoted by subscripts 1 and 2, the free energy of binding is given by $\Delta G = e_1 + e_2 + c + p$. Conformational free energies, $c$, account for changes in DNA and repressor conformation, which are needed to accommodate multiple simultaneous interactions (Figure 2).

All these contributions to the free energy, taking into account the three operators for specific binding of the *lac* repressor, can be expressed in mathematical terms as

$$\begin{aligned}\Delta G(s) = &(p+e_1)s_1 + (p+e_2)s_2 + (p+e_3)s_3 \\ &+ (c_{L12} - ps_1s_2)s_{L12} + (c_{L13} - ps_1s_3)s_{L13} + (c_{L23} - ps_2s_3)s_{L23} \\ &+ \infty(s_{L12}s_{L13} + s_{L12}s_{L23} + s_{L13}s_{L23}),\end{aligned} \quad (4)$$

where $s_1$, $s_2$, and $s_3$ are state variables that can take the values 0 and 1 to indicate whether ($=1$) or not ($=0$) the repressor is bound to $O_1$, $O_2$, and $O_3$, respectively; and $s_{L12}$, $s_{L13}$, and $s_{L23}$ are variables that indicate whether ($=1$) or not ($=0$) DNA forms the loops $O_1$-$O_2$, $O_1$-$O_3$, and $O_2$-$O_3$, respectively. The subscripts of the different contributions to the free energy have the same meaning as those of the corresponding binary variables. The infinity in the last term of the free energy implements that two loops that share one operator cannot be present simultaneously by assigning an infinite free energy to those states (18).

The set of 6 state variables, denoted by $s = (s_1, s_2, s_3, s_{L12}, s_{L13}, s_{L23})$, describes the specific binding configuration of the repressor-DNA complex. For instance, a repressor bound to $O_2$ is specified by $s = (0,1,0,0,0,0)$; a repressor bound to $O_1$ and $O_3$ looping the intervening DNA, by $s = (1,0,1,0,1,0)$; and three repressors bound, one to each operator, by $s = (1,1,1,0,0,0)$. The specific value of the free energy is obtained by substituting the values of the state variables in the expression of the free energy. This description in terms of state variables can be visualized as a factor graph (Figure 3).

The probability of any of these states depends exponentially on its free energy and is obtained from statistical thermodynamics as

$$P_s = \frac{e^{-\Delta G(s)/RT}}{Z}, \tag{5}$$

where $RT$ is the gas constant times the absolute temperature. The partition function, $Z = \sum_s e^{-\Delta G(s)/RT}$, is used as normalization factor.

## Transcriptional control

Gene expression in the *lac* operon is completely abolished when the repressor is bound to $O_1$; otherwise, transcription takes place either at an activated maximum rate, $\Gamma_{max}$, when $O_3$ is free or at a basal reduced rate, $\chi \Gamma_{max}$, when $O_3$ is occupied. This reduction by a factor $\chi$ arises because binding of the repressor to $O_3$ prevents CAP from activating transcription (13, 18).

The transcription rate $\Gamma(s)$ can be expressed in terms of state variables as

$$\Gamma(s) = \Gamma_{max}(1-s_1)(\chi s_3 + (1-s_3)). \tag{6}$$

With this approach, the effective transcription rate,

$$\bar{\Gamma} = \sum_s \Gamma(s) P_S = \frac{1}{Z} \sum_s \Gamma(s) e^{-\Delta G(s)/RT} \tag{7}$$

is obtained by computing the thermodynamic average over all the representative states; namely, by performing the sum above over all possible combination of values of *s*.

## Model calibration

The overall model has only two free parameters: the constants *a* and *b* that relate scores to free energies of binding. Their values are inferred by minimizing the square logarithmic error between measured and model normalized transcription ($\bar{\Gamma}/\Gamma_{max}$). The values of the other four parameters, three conformational free energies and CAP activation, are taken from the experimental data. Explicitly, the value $\chi$=0.03 was reported in Ref. (26); the value $c_{L12}$=23.35 kcal/mol was obtained in Ref. (20) from experimental data in Ref. (26); the values $c_{L13}$=22.05 kcal/mol and $c_{L23}$=23.50 kcal/mol were obtained from the value of $c_{L12}$ by taking into account the dependence of the conformational free energy on the distance between operators (20, 27, 28) and the stabilization of the $O_1$-$O_3$ loop by CAP (29, 30).

# Results and Discussion

We applied the multisite approach to classic experiments on the *lac* operon that considered gene expression for different repressor concentrations in *E. coli* strains covering all eight possible combinations of operator deletions (14). The sequences of the three wild-type (WT) operators $O_1$, $O_2$, and $O_3$ were used to compute the PWM from which we obtained the scores for these three operators and their respective "deletions" $O_{1M}$, $O_{2M}$, and $O_{3M}$ (See Table 1). The scores correctly ranked the three WT operators according to their measured strength and consistently ranked all the deletions below all the WT operators.

The values of the parameters *a* and *b* were obtained by fitting the model to the experimental transcription data using

$$\Delta G(s) = (p + aS_1 + b)s_1 + (p + aS_2 + b)s_2 + (p + aS_3 + b)s_3 \\ + (c_{L12} - ps_1s_2)s_{L12} + (c_{L13} - ps_1s_3)s_{L13} + (c_{L23} - ps_2s_3)s_{L23} \\ + \infty(s_{L12}s_{L13} + s_{L12}s_{L23} + s_{L13}s_{L23}) \tag{8}$$

as the free energy of the system. This expression is obtained after substitution of the relation $e = aS + b$ in Equation 4. In this way, the binding is described by the PWM scores $S_1$, $S_2$, and $S_3$ for each site together with the conformational contributions to the free energy from DNA looping (28).

The model, with just $a$ and $b$ as free parameters, is able to fit the experimental data (14) within 0.29-fold accuracy over a 10,000-fold range of transcriptional activity (Figure 4A). In total, there are 22 experimental points, accounting for 8 operator configurations, three different repressor concentrations, and different functional forms of the transcription curves. The value $FA$ that quantifies the ability of the model to capture the experimental data within $FA$-fold accuracy is explicitly defined for a set of $N$ experimental, $\Gamma ex$, and computed, $\Gamma cp$, transcription rates through the expression $N\log(1+FA)^2 = \sum_{i=1}^{N}\log(\Gamma ex_i/\Gamma cp_i)^2$, and it indicates that typically measured and computed values differ from each other by a factor $1+FA$.

The interaction free energies obtained from the model for the best-fit $a$ and $b$ parameters and the corresponding experimental *in vivo* values (18) are shown in Table 1. The results of the model exhibit a good agreement with the available experimental data. In terms of dissociation constants, the differences between the predicted and observed values are within the 2-fold range (Table 1). An advantage of the approach we have followed is that the *in vivo* free energies, and the corresponding dissociation constants, take into account implicitly the effects of non-specific binding. The reason is that their values are measured with respect to the reference state with no repressor bound to the operators, which includes the repressors in solution in the cytosol as well as the repressors bound non-specifically to DNA (for a detailed quantitative discussion see Appendix II of Ref. (16)).

To test the predictive potential of the multisite model, we used experimental data sets for two operator configurations to infer the values of the parameters $a$ and $b$, and then used the calibrated model to predict the transcriptional activity for the other six configurations (Figure 4B). The model accuracy at predicting new data decreases only slightly with respect to the all-fit accuracy. In principle, only two experimental data points would be needed to calibrate model because there are only two free parameters. Indeed, just two experimental points can be used to calibrate the model with just a slight additional decease in global accuracy (Figure 4C). Therefore, without using any free energy of binding, the multisite model is able to accurately predict gene expression curves over a 10,000-fold range for eight different *E. coli* strains covering all possible combinations of operator "deletions" from just two experimental calibration data points and the sequences of the six DNA sites involved.

There is an important prediction that goes beyond the experimentally observed free energies of binding. The deletion $O_{1M}$ of the main operator $O_1$ involved the mutation of just three DNA base pairs. As a consequence, the model predicts for $O_{1M}$ an increase in free energy of 5.4 kcal/mol with respect to $O_1$, or equivalently about an 8,000-fold increase of the dissociation constant, which is substantial but still remains relatively close to the free energy of binding to $O_3$, the weakest WT operator (Table 1). We found that such a decrease has transcriptional consequences that make it distinguishable from a complete deletion (Figure 5). Thus, the multisite approach is able not only to both accurately predict gene expression and recover known free energies but also to obtain precise affinity estimates for very weak sites that were assumed not to bind the *lac* repressor.

Typically, the effects of a given sequence depend on the context. This dependence has been noted explicitly as one of the main limiting factors for identifying physiologically relevant sites and for linking statistical sequence information, such as PWM scores, to transcriptional activity (31). This fundamental problem in gene regulation is believed to result from the interplay among multiple DNA sites in orchestrating the binding patterns of transcription factors that control gene expression (2). The approach presented here overcomes this limitation by using detailed biophysical modeling of multidomain binding to directly connect statistical sequence information with transcriptional activity. We have shown that, for the prototypical *lac* operon, which relies on a cluster of three non-adjacent sites over a 0.5 kb DNA region to control transcription, this multisite approach accurately recapitulates the observed transcriptional activity over a 10,000-fold range for all the possible combinations of operator deletions.

## Acknowledgments

This work was supported by the MICINN under grant FIS2009-10352.

# Tables

**Table 1: Operator sequences and their statistical and binding properties**

| Name | Sequence | $S$ | $aS+b$ (kcal/mol) | $e$ (kcal/mol) | $K_D^{sc}$ (nM) | $K_D^{ex}$ (nM) |
|---|---|---|---|---|---|---|
| $O_1$ | AATTGTGAGCGGATAACAATT | -13.38 | -27.62 | -27.8 | 0.728 | 0.54 |
| $O_2$ | AAATGTGAGCGAGTAACAACC | -12.17 | -25.94 | -26.3 | 12.1 | 6.62 |
| $O_3$ | GGCAGTGAGCGCAACGCAATT | -10.95 | -24.25 | -24.1 | 201 | 259 |
| $O_{1M}$ | AATTGT<u>T</u>AGCGGA<u>G</u>AA<u>G</u>AATT | -9.51 | -22.26 | N/A | 5600 | N/A |
| $O_{2M}$ | <u>G</u>AA<u>G</u>GT<u>T</u>A<u>A</u>TGA<u>A</u>TA<u>G</u>CA<u>C</u>CC | -5.12 | -16.16 | N/A | $1.44\times10^8$ | N/A |
| $O_{3M}$ | <u>TCG</u>A<u>TC</u>GAGC<u>T</u>CAACGCAATT | -4.71 | -15.60 | N/A | $3.37\times10^8$ | N/A |

The PWM score, $S$, for a given operator sequence is used to estimate its interaction free energy with the *lac* repressor as $aS+b$, with $a=1.387$ kcal/mol and $b=-9.064$ kcal/mol. The experimental values of these free energies, $e$, are from Ref. (18). Dissociation constants are computed as $K_D^{sc}=e^{(aS+b+p°)/RT}$ for the predictions from PWM scores and as $K_D^{ex}=e^{(e+p°)/RT}$ for the experimental data. N/A stands for data not available.

# Figure legends

**Figure 1:** Integration of sequence statistics into predictive biophysical multidomain models. The approach is implemented by considering first the three operators as DNA signals. They are used to construct a probabilistic model that provides binding scores for these and similar mutated sequences. The scores are subsequently linked parametrically to binding free energies and incorporated directly into a detailed biophysical model of transcription regulation. The link between scores and free energies is calibrated by fitting the model to a subset of experimental data. The calibrated model is then tested with different sets of data.

**Figure 2:** Operator locations on DNA and binding of the *lac* repressor. **(A)** The main, $O_1$, and the two auxiliary, $O_2$ and $O_3$, operators are shown as black rectangles on the black line representing DNA. Binding of the *lac* repressor to $O_1$ prevents transcription of the three lacZYA genes. **(B)** A repressor is shown bound to the operator $O_2$. The free energy of binding is $\Delta G = e_2 + p$. **(C)** A repressor is shown looping DNA by binding simultaneously to $O_1$ and $O_3$. The free energy of this binding configuration is $\Delta G = e_1 + e_3 + c_{L13} + p$.

**Figure 3:** Factor graph for the free energy components of the multisite *lac* represor-operator binding. The free energy of the system, $\Delta G(s)$, as a function of the state variables, $s = (s_1, s_2, s_3, s_{L12}, s_{L13}, s_{L23})$, has a graphical representation in the form of factor graph. The round nodes represent state variables and the rectangular nodes represent contributions to the free energy. The quantity in the rectangular node is present in the free energy when all its connecting state variables are equal to 1. The experimental values for wild-type parameters are $e_1$=-27.8 kcal/mol, $e_2$=-26.3 kcal/mol, $e_3$=-24.1 kcal/mol, $c_{L12}$=23.35 kcal/mol, $c_{L13}$=22.05 kcal/mol, and $c_{L23}$ =23.50 kcal/mol. The dependence on the *lac* repressor concentration, *n*, is given by the positional free energy, $p=p°-RT \ln n$, with $p°$=15 kcal/mol.

**Figure 4:** Model calibration and prediction of the transcriptional activity as a function of the repressor concentration. The normalized transcription ($\bar{\Gamma}/\Gamma_{max}$) was obtained for WT and seven mutants accounting for all the combinations of deletions of the three operators. For each of the eight cases, the results of the model (continuous lines) as a function of the repressor concentration are compared with the experimental data from Ref. (14) (square symbols). The particular set of WT or "deleted" operators is indicated for each curve; for instance, $O_1$-$O_2$-$O_3$ corresponds to WT *lac* operon and $O_{1M}$-$O_{2M}$-$O_{3M}$, to the mutant with all three operators "deleted". The values of the experimental parameters used are $c_{L12}$=23.35 kcal/mol, $c_{L13}$=22.05 kcal/mol, $c_{L23}$ =23.50 kcal/mol, and $\chi$=0.03. The PWM scores, $S$, for each site are as shown in Table 1. **(A)** The values of the parameters $a = 1.387$ kcal/mol and $b = -9.064$ kcal/mol that connect interaction free energies with scores, $e = aS + b$, were obtained by fitting the model to all the experimental transcription data. **(B)** The values of the parameters $a = 1.348$ kcal/mol and $b = -9.531$ kcal/mol were obtained by fitting the model to the experimental data for the operator configurations $O_1$-$O_2$-$O_3$ and $O_{1M}$-$O_2$-$O_3$. The model accurately predicts the normalized transcription for the other six operator configurations. **(C)** Only two experimental points (indicated by big gray circles) are used to obtain the values of the parameters $a = 1.462$ kcal/mol and $b = -8.208$ kcal/mol. The model is still able to accurately predict the normalized transcription for the remaining 20 experimental points.

**Figure 5:** Deletions *versus* weak binding. The normalized transcription ($\bar{\Gamma}/\Gamma_{max}$) for the four configurations with $O_{1M}$ is shown for the model as in Figure 4A (continuous line), for the model assuming that the free energy of binding to $O_{1M}$ is infinite as in a complete deletion (discontinuous line), and for the experimental data from Ref. (14) (square symbols).

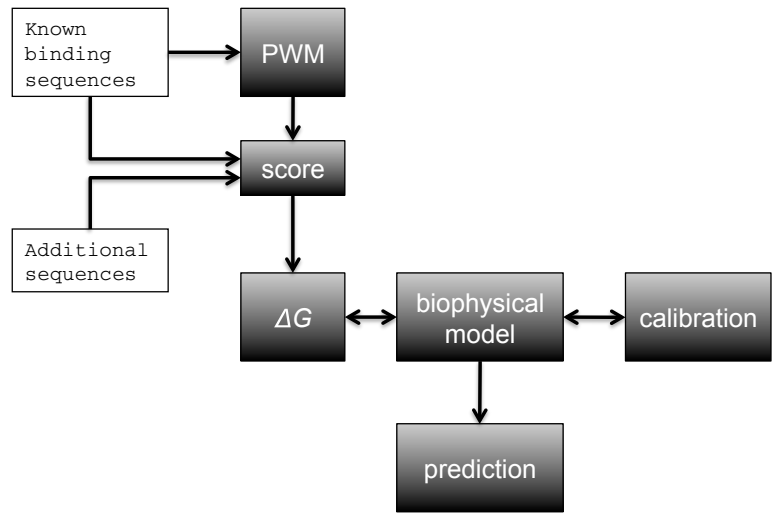

Figure 1

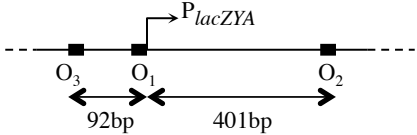
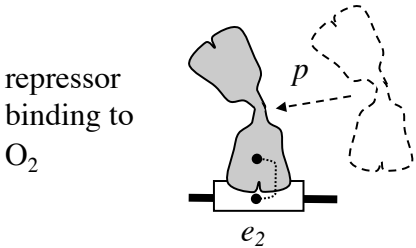
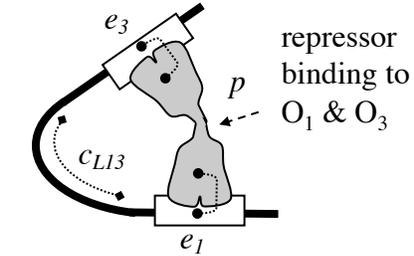

Figure 2

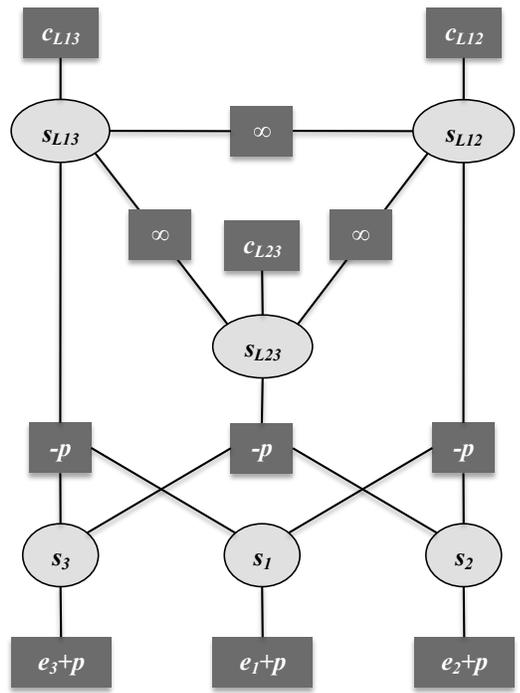

Figure 3

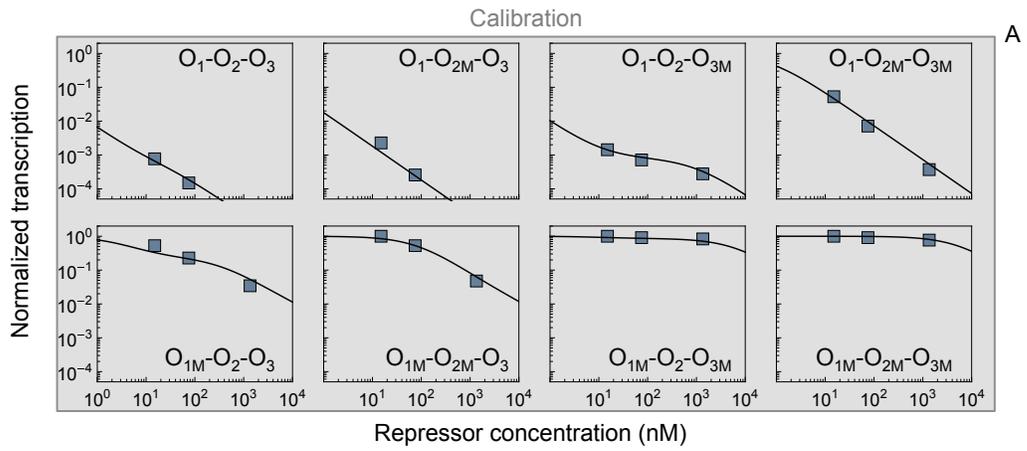
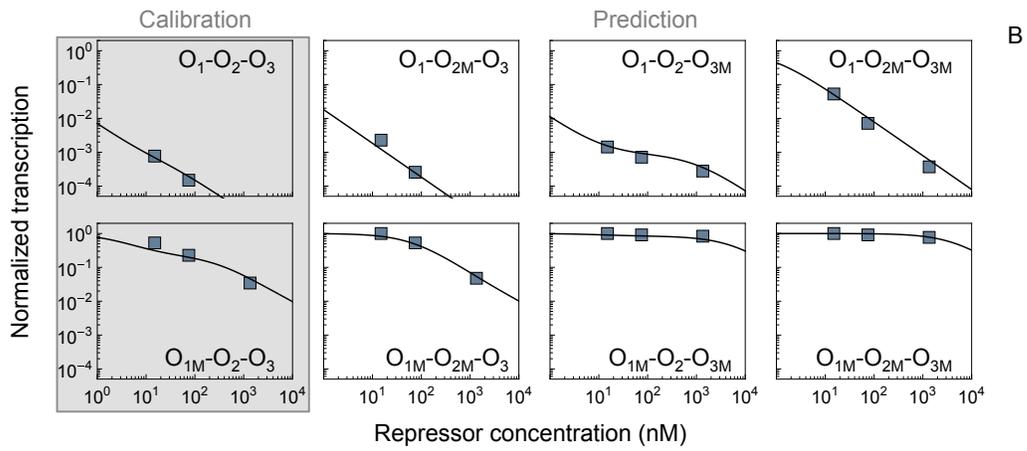
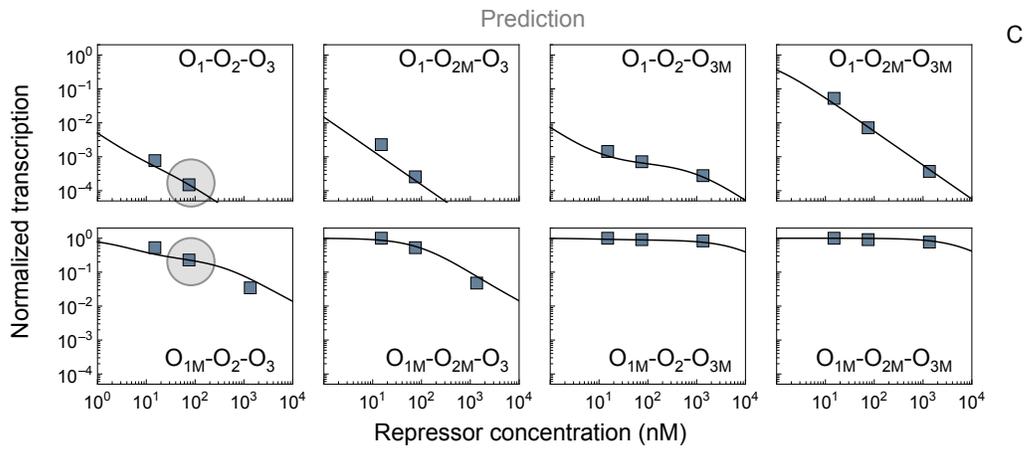

Figure 4

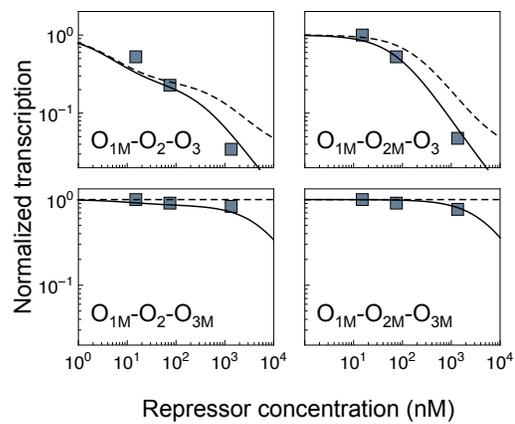

Figure 5